\documentclass[10pt,aps,prd,reprint,nofootinbib]{revtex4-1}

\usepackage{amsmath}
\usepackage{amssymb}
\usepackage{fontawesome}
\usepackage{graphicx}
\usepackage[colorlinks,pdfusetitle]{hyperref}
\hypersetup{colorlinks=true,allcolors=[rgb]{1,0.56,0}}

\newcommand{\orcid}[1]{\href{https://orcid.org/#1}{#1}}
\newcommand{\pdn}[3]{\frac{\partial^{#1}#2}{\partial #3^{#1}}}

\begin{document}

\title{Techniques for Solving Static Klein-Gordon Equation with Self-Interaction $\lambda\phi^4$ and Arbitrary Spherical Source Terms}

\author{Peter B.~Denton}
\email{pdenton@bnl.gov}
\thanks{\orcid{0000-0002-5209-872X}}
\affiliation{High Energy Theory Group, Physics Department, Brookhaven National Laboratory, Upton, NY 11973, USA}

\begin{abstract}
The Klein-Gordon equation for a scalar field sourced by a static spherically symmetric background is an interesting second-order differential equation with applications in particle physics, astrophysics, and elsewhere.
Here we present static solutions for generic source density profiles in the case where the scalar field has no interactions or a mass term.
For a $\lambda\phi^4$ self-interaction term, we develop the techniques that are necessary numerical computation.
We also provide code to perform the numerical calculations that can be adapted for arbitrary density profiles.
\href{https://github.com/PeterDenton/Spherically-Symmetric-SelfInteracting-Scalar-Solver}{\large\faGithub}
\end{abstract}

\date{July 15, 2024}

\maketitle

\section{Introduction}
The Klein-Gordon equation allows one to compute the behavior of a relativistic field $\phi(t,x,y,z)$.
The massless form with no interactions is
\begin{equation}
\square\phi=0\,,
\end{equation}
where $\square$ is the d'Alembert operator $\square=\pdn2{}t-\pdn2{}x-\pdn2{}y-\pdn2{}z$.
The more general version of the Klein-Gordon equation includes a mass term for $\phi$ as well as additional interactions.
\begin{equation}
\square\phi+V'(\phi)=0\,,
\end{equation}
where $V$ is the potential and the prime denotes the derivative with respect to the field.
The Klein-Gordon equation \cite{1926_klein,1926_gordon,1926_fock} has been extensively investigated throughout the math and physics literature in a variety of environments, see e.g.~\cite{Frasca:2005sx,Frasca:2009bc,Natario:2019sap,2020arXiv200615688G,2020arXiv200801276K,2021arXiv210914902B,Taylor:2022wah,2022arXiv220411261S,2023ArRMA.247...17G} for an incomplete list.
The potential for a massive field that is also self interacting is
\begin{equation}
V(\phi)=\frac12m^2\phi^2+\frac1{4!}\lambda\phi^4\,,
\label{eq:V}
\end{equation}
where $m$ is the mass of the field and $\lambda$ is a dimensionless parameter of the model indicating the strength of the interaction.
Such a potential has been considered in many different physical environments including loops, renormalization effects, sources, and perturbative solutions among others, see e.g.~\cite{Coleman:1973jx,Dvali:2011uu,Yoda:2012xu,Shi:2021inw,Davoudiasl:2023uiq}.
Note that a $\phi^4$ term may be induced at the loop level or may contribute to loop calculations.
We ignore these issues here and treat $\phi$ as a classical field which is relevant when $\lambda\ll1$.
In addition, one could also consider a cubic term $\propto\phi^3$ which we only briefly discuss in section \ref{sec:cubic} below and otherwise ignore as it turns out the $\lambda\phi^4$ term will contain most of the challenges of the more general potential.
From a physics perspective a cubic term will generically lead to non-zero vacuum expectation values which could be desirable in some environments, and be undesirable in others.
Finally, we note that we are fixing the signs of the terms in eq.~\ref{eq:V}; the Higgs field is a well studied example with a different sign on the $m^2\phi^2$ term.
The numerical approach is equally valid in either case, while care is required to extend the analytic results.

The final contribution to the Klein-Gordon equation we consider is a source term $n(r)\ge0$ where $n(r)$ is some background field charge density profile that is coupled to $\phi$ via an additional interaction.
We also assume that $n(r)=0$ for $r>r_s$ where $r_s$ is the finite radius of the object, although one could take $r_s\to\infty$ as well.
Note that we have taken $n$ to be spherically symmetric and depend only on the radius and not direction or time.
We will also assume that $\lim_{r\to0}n(r)$ is finite.
These conditions are all met for celestial objects such as planets or stars.
Now the full Klein-Gordon equation we will consider is
\begin{equation}
\square\phi+m^2\phi+\frac1{3!}\lambda\phi^3+n(r)=0\,.
\end{equation}

Since the equation is independent of time or direction in spherical coordinates, we make the ansatz that $\phi(t,x,y,z)=\phi(r)$.
This is essentially the statement that we are near the source and that the source makes a significant contribution to the field such that any time dependence can be safely ignored.
Then, in spherical coordinates, we have the non-linear second order differential equation we wish to solve:
\begin{equation}
-2\frac{\phi'}r-\phi''+m^2\phi+\frac1{3!}\lambda\phi^3+n(r)=0\,,
\label{eq:KG24 spherical}
\end{equation}
where the primes denote derivatives with respect to $r$.

Before we begin to solve it, we note several features.
\begin{itemize}
\item Being a second order differential equation, two initial conditions are necessary (but not necessarily sufficient) to uniquely determine the field.
The initial conditions will depend on the underlying physics of the problem, outlined here.
\item The first of the initial conditions can be determined at the center of the sphere where symmetry arguments find that we must have $\phi'(0)=0$, provided that $\phi$ is $C^1$ through the origin.
Diverging derivatives would require that the field diverges so this possibility can be discarded, provided that a finite solution can be found.
\item Next, we have that the field should satisfy $\phi(\infty)=0$ since in the absence of a source term we want the field to return to the vacuum state.
\item The inclusion of the $\lambda$ term makes the equation non-linear.
Thus one cannot use superposition to take a single solution and form up the total solution as one can in the case where $\lambda=0$.
\end{itemize}

Before addressing the full solution in its generality, we solve it in simpler cases and learn useful strategies for the full solution.

\section{Massless Non-Self Interacting Case}
We begin with the simplest possible case: $m=\lambda=0$.
The differential equation is
\begin{equation}
-2\frac{\phi'}r-\phi''+n(r)=0\,.
\label{eq:KG spherical}
\end{equation}
This is equivalent to electrodynamics and can be solved in many ways.
We focus on deriving a general solution to the differential equation.

The solution for $r>r_s$ is
\begin{equation}
\phi(r)=\frac Ar+B\,,
\label{eq:phi AB}
\end{equation}
for free parameters $A$ and $B$.
We see that $B=0$ since $\phi(\infty)=0$.

Inside the sphere the general solution is
\begin{equation}
\phi(r)=C+\int_0^r\left(\frac D{r_1^2}+\frac g{r_1^2}\int_0^{r_1}n(r_2)r_2^2dr_2\right)dr_1\,,
\label{eq:phi CD}
\end{equation}
for some free parameters $C$ and $D$.
From the initial condition $\phi'(0)=0$ we must have $D=0$.

It is useful to know $\phi(0)$ for numerical purposes.
We calculate $\lim_{r\to0}\phi(r)$ directly from the differential equation, eq.~\ref{eq:KG spherical}, by multiplying by $r$ (this does not create a problem for evaluation as $r\to0$ since $\phi\in C^1$) and integrating over $r$:
\begin{equation}
-2\int_0^\infty\phi'dr-\int_0^\infty r\phi''dr+\int_0^{r_s}n(r)rdr=0\,.
\end{equation}
We now use the fact that $\lim_{r\to\infty}r\phi'(r)=0$ since $\phi'(r)$ must fall off faster than $1/r$ as $r\to\infty$ since if $\phi'(r)\to1/r$ then $\phi(r)\to\log(r)$ at large $r$ diverges which fails our second initial condition requirement.
We also use $\phi'(0)=0$ and find that
\begin{equation}
\phi(0)=-\int_0^{r_s}rn(r)dr\,.
\label{eq:KG phi0}
\end{equation}
This allows for a determination of $C$ in eq.~\ref{eq:phi CD}.
Combining eq.~\ref{eq:KG phi0} with eq.~\ref{eq:phi CD} for $r<r_s$, we have
\begin{equation}
\phi(r)=\int_0^r\left(\frac1{r_1^2}\int_0^{r_1}n(r_2)r_2^2dr_2\right)dr_1-\int_0^{r_s}n(r_1)r_1dr_1\,.
\end{equation}
Finally we determine $A$ in eq.~\ref{eq:phi AB} to be
\begin{equation}
A=-gr_s\int_0^{r_s}\left(n(r)r-\frac1{r^2}\int_0^rn(r_1)r_1^2dr_1\right)dr\,,
\end{equation}
completing the calculation.

\subsection{An Example}
A common example is to consider a uniform hard sphere of charge given by
\begin{equation}
n(r)=
\begin{cases}
n\qquad&r<r_s\\
0&r>r_s
\end{cases}\,.
\label{eq:nr hard}
\end{equation}
While astrophysical bodies typically have $n'(r)\lneq0$ for $r<r_s$, this approximation may still be useful for showing the behavior of the field from a large core sourcing most of the field strength.

We show that these equations reproduce the known behavior.
First we find that
\begin{equation}
A=-\frac13nr_s^3\,.
\end{equation}
We define the total charge
\begin{equation}
Q=4\pi\int_0^{r_s}n(r)r^2dr=\frac43\pi nr_s^3\,,\label{eq:Q}
\end{equation}
where for the second equation we have set $n(r)$ to the step function in eq.~\ref{eq:nr hard}.
Then the solution reproduces the familiar result
\begin{equation}
\phi(r)=
\begin{cases}
-\frac Q{8\pi r_s}\left(3-\frac{r^2}{r_s^2}\right)\qquad&r<r_s\\
-\frac Q{4\pi r}&r>r_s
\end{cases}\,.
\end{equation}

\section{Massive Non-Self Interacting Case}
We now consider $m\neq0$ but still keep $\lambda=0$.
This is often referred to as a Yukawa potential \cite{Yukawa:1935xg}.
The differential equation is
\begin{equation}
-2\frac{\phi'}r-\phi''+m^2\phi+n(r)=0\,,
\label{eq:KG2 spherical}
\end{equation}
which has also been studied in a variety of contexts including with the spherical source term \cite{Yukawa:1935xg,Smirnov:2019cae,Acevedo:2023owd}.

For $r>r_s$ the solution is
\begin{equation}
\phi(r)=\frac Are^{-mr}+\frac Bre^{mr}\,,
\end{equation}
for free parameters $A$ and $B$.
Again, $B=0$ since $\phi(\infty)=0$ and without loss of generality we have taken $m>0$.

Inside the sphere the solution is of the form
\begin{equation}
\phi(r)=\frac Cre^{-mr}+\frac Dre^{mr}+I(r)\,,
\end{equation}
where
\begin{align}
I(r)={}&\frac g{2mr}\left[e^{mr}\int_0^re^{-mr_1}n(r_1)r_1dr_1\right.\nonumber\\
&\left.-e^{-mr}\int_0^re^{mr_1}n(r_1)r_1dr_1\right]\,,\\
I'(r)={}&\frac{-g}{2mr^2}\left[e^{mr}(1-mr)\int_0^re^{-mr_1}n(r_1)r_1dr_1\right.\nonumber\\
&\left.-e^{-mr}(1+mr)\int_0^re^{mr_1}n(r_1)r_1dr_1\right]\,.
\end{align}
Note that $\lim_{r\to0}I'(r)=0$ so long as $n(r)$ is finite at the origin.
To satisfy the initial condition $\phi'(0)=0$ we must have $C=-D$.
Then we write
\begin{equation}
\phi(r)=2\frac Dr\sinh(mr)+I(r)\,.
\label{eq:phi KG2 small r}
\end{equation}

We now require that $\phi$ and $\phi'$ are continuous at $r=r_s$.
We find
\begin{align}
A={}&I(r_s)r_s\cosh(mr_s)-\frac1m\left[I(r_s)+I'(r_s)r_s\right]\sinh(mr_s)\,,\\
D={}&-\frac{e^{-mr_s}}{2m}\left[I(r_s)(1+mr_s)+I'(r_s)r_s\right]\,.
\label{eq:D KG2}
\end{align}
Thus $I(r)$ and $I'(r)$ contain all the non-trivial information about $n(r)$ necessary to compute $\phi(r)$ anywhere.
This completes the calculation.

\subsection{An Example}
\label{sec:yukawa example}
We again repeat the example of the previous section where $n(r)$ is a hard sphere as defined in eq.~\ref{eq:nr hard}.
Then we have
\begin{align}
I(r)&=\frac n{m^3r}\left[\sinh(mr)-mr\right]\,,\\
I'(r)&=-\frac n{m^3r^2}\left[\sinh(mr)-mr\cosh(mr)\right]\,.
\end{align}
This leads to coefficients of
\begin{align}
A&=\frac n{m^3}\left[\sinh(mr_s)-mr_s\cosh(mr_s)\right]\,,\\
D&=-\frac n{2m^3}\left[1-e^{-mr_s}-mr_se^{-mr_s}\right]\,.
\end{align}
Thus we write the complete solution for the hard sphere of charge as
\begin{equation}
\phi(r)=
\begin{cases}
-\frac n{m^2}\left[1-e^{-mr_s}(1+mr_s)\frac{\sinh(mr)}{mr}\right]&r<r_s\\
-\frac n{m^2}\left[mr_s\cosh(mr_s)-\sinh(mr_s)\right]\frac{e^{-mr}}{mr}&r>r_s
\end{cases}\,.
\end{equation}
It is easy to verify that this recovers the massless solution in the limit $m\to0$.

Analytic expressions also exist for some other density profiles such an exponential density function of the form $n(r)=n_0e^{-r\kappa}$, see e.g.~\cite{Smirnov:2019cae}.

\subsection{Beyond a point charge}
The above solution presented in section \ref{sec:yukawa example} is for a hard sphere of charge, we now elaborate some of its properties.
First, in the limit that $r_s$ is small compared to $1/m$, but $n/r_s^3$ proportional to the total charge $Q$ defined in eq.~\ref{eq:Q} is held constant, we find that we recover the expected Yukawa potential expression valid for $r\gg r_s$
\begin{equation}
\lim_{r_s\to0}\phi(r)=-\frac Q{4\pi}\frac{e^{-mr}}r\,.
\end{equation}
In the case where $r_s$ is finite, however, there are corrections to the above formula and the solution takes on a different functional form than if it were sourced by a delta function.
To see this difference, we expand the solution in small $mr_s$ and the next several terms in the $mr_s\ll1$ expansion are
\begin{align}
\phi(r)={}&-\frac Q{4\pi}\frac{e^{-mr}}r\left\{1\vphantom{\frac12}\right.\nonumber\\
&\left.+\frac1{10}\left(mr_s\right)^2+\frac1{280}\left(mr_s\right)^4+\mathcal O[(mr_s)^6]\right\}\,.
\label{eq:non point}
\end{align}
Thus a Yukawa potential for a massive mediator depends not only on the total charge of the source, but also the size of the source.
Note that linearity applies here since $\lambda=0$ so a non-uniform source can be summed or integrated.

We see from eq.~\ref{eq:non point} that if $mr_s\sim1$ then there is a $\sim10\%$ correction to the point charge assumption, and it takes only $mr_s\sim3$ to get a factor of 2 correction from the point charge approximation.
Objects with a declining density profile $n'(r)<0$ will see a smaller correction as they are closer to a point charge.

\subsection{Surface to Core Ratio}
A quantity that may be of interest for some problems is the ratio of the field strength at the surface of the sphere to the center.
Continuing with $\lambda=0$, for limiting values of $m$ we find that the ratio is
\begin{equation}
R_{sc}\equiv\frac{\phi(r_s)}{\phi(0)}=
\begin{cases}
\dfrac{\int_0^{r_s}r^2n(r)dr}{r_s\int_0^{r_s}rn(r)dr}\quad&m=0\\[0.2in]
\dfrac12\dfrac{n(r_s)}{n(0)}&m\to\infty
\end{cases}\,,
\end{equation}
where the $\frac12$ comes from assuming that we are exactly at $r=r_s$.
In the hard sphere case this simplifies to
\begin{equation}
R_{sc}=\frac{\sinh(mr_s)-mr_s\cosh(mr_s)}{mr_s-e^{mr_s}+(mr_s)^2}\,,
\end{equation}
which is, in the limits,
\begin{equation}
R_{sc}=
\begin{cases}
\frac23\quad&m=0\\[0.1in]
\frac12&m\to\infty
\end{cases}\,,
\end{equation}
but has a local minimum at $mr_s\simeq3.93$ where $R_{sc}\simeq0.41$.
So the allowed range of $R_{sc}$ for any value of $m$ but with $\lambda=0$ is $[0.41,\frac23]$.

We note that so long as $n'(r)<0$ (which is what we expect in celestial situations), then $R_{sc}\lneq\frac23$ at $m=0$.
Similarly, in the $m\to\infty$ limit $R_{sc}\lneq\frac12$.

\section{Massive Self Interacting Case}
We now address the solution of the full differential equation from eq.~\ref{eq:KG24 spherical} including the $\lambda\phi^3$ term.
We are not aware of a general analytic solution to this equation.
Ref.~\cite{Frasca:2009bc} determined the time and space varying wave solution in the absence of a source term.
Refs.~\cite{Burrage:2021nys,Braden:2020zfa} investigated this equation with a source term but with the opposite sign on $m^2$.
To understand the behavior of the full equation we turn to approximations and numerical solutions.

We begin by identifying several useful properties of this equation to aid in numerical solutions.
First, we note that the term $\phi'/r$ becomes $0/0$ as $r\to0$.
In the limit $r\to0$, since $\phi'(0)=0$, we have that $\phi''(0)=\lim_{r\to0}\phi'(r)/r$ by L'H\^opital's rule.
Thus
\begin{equation}
\lim_{r\to0}\frac{\phi'(r)}r=\frac13\left[m^2\phi(0)+\frac\lambda{3!}\phi^3(0)+n(0)\right]\,.
\label{eq:x}
\end{equation}
So all that is needed to determine this ratio is $\phi(0)$.
Moreover, $\phi(0)$ is also needed to have both initial conditions at the origin (the other being $\phi'(0)=0$) and allows us to calculate eq.~\ref{eq:x} to evaluate the differential equation near the origin.

The value of $\phi(0)$ is ultimately set from the fact that $\phi(\infty)=0$.
In principle, one can numerically vary $\phi(0)$ until one finds that $\phi(\infty)\to0$; that is, one can use a shooting method to convert this boundary value problem to an initial value problem.
This strategy is numerically viable by following a procedure that maximizes the distance $r$ at which either $\phi(r)$ crosses zero or when the derivative changes from positive to negative.

This strategy suffers from two main problems.
First, a sufficiently accurate initial guess is needed to realistically converge on the correct value of $\phi(0)$.
Second, when $\lambda$ is large enough to contribute, the differential equation may become quite stiff depending on the shape of the density profile.
For example, for some interesting parameters, we find that changing the initial condition $\phi(0)$ by one part in the $10^{14}$, close to the double precision limit, the numerical solution varies between diverging to positive infinity and negative infinity well within $r_s$.
Thus we find that while it may be possible to evaluate $\phi(0)$ to good precision, it is not always sufficient to fully evolve the function to any value of $r$ desired.
Arbitrary precision techniques are likely required here.

\begin{figure}
\centering
\includegraphics[width=\columnwidth]{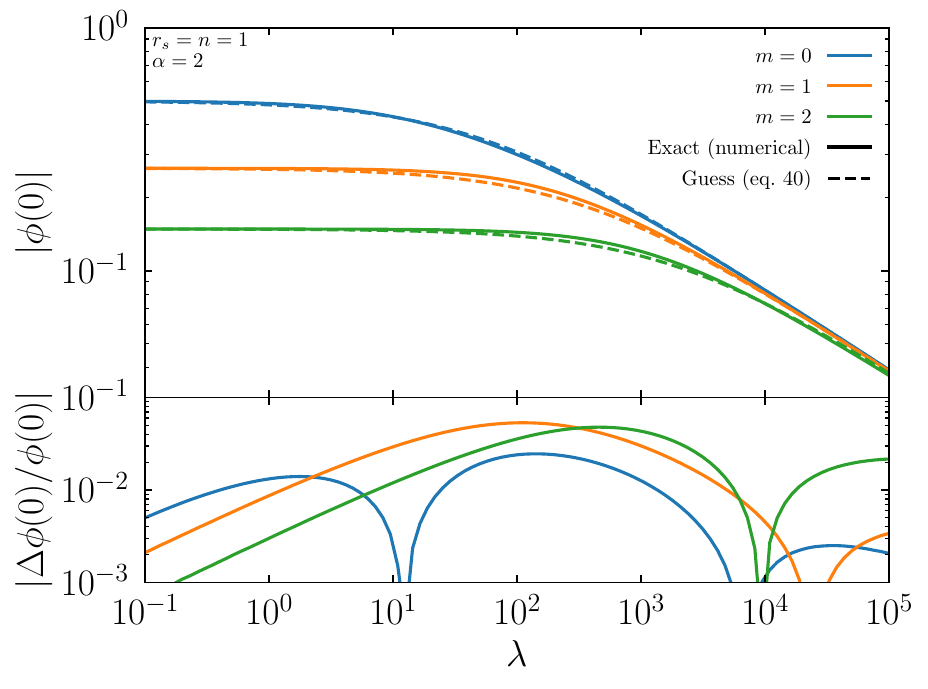}
\caption{\textbf{Top}: The numerical value of $|\phi(0)|$ as a function of $\lambda$ for different representative values of $m$.
The dashed curves are the approximations using eq.~\ref{eq:phi0 approximation}.
We use a hard sphere of charge with $n=1$ and $r_s=1$ and fix $\alpha=2$ for the approximation.
\textbf{Bottom}: The fractional error between the exact numerical and approximate calculations.}
\label{fig:phi0}
\end{figure}

To aid with the first issue, we found some expressions that serve well to approximate $\phi(0)$ by examining its behavior in different limits and then connecting them together.
In the small $\lambda$ limit we recover the solution from eqs.~\ref{eq:phi KG2 small r} and \ref{eq:D KG2} which is
\begin{equation}
\lim_{\lambda\to0}\phi(0)=-e^{-mr_s}\left[I(r_s)(1+mr_s)+I'(r_s)r_s\right]\equiv f_0
\end{equation}
In the hard sphere case, this becomes
\begin{equation}
\lim_{\lambda\to0}\phi(0)=-\frac n{m^2}\left[1-e^{-mr_s}(1+mr_s)\right]\,.
\end{equation}
In the massless case this returns the known result,
\begin{equation}
\lim_{\lambda,m\to0}\phi(0)=-\frac{nr_s^2}2\,.
\end{equation}
Second, we consider $\phi(0)$ in the large $\lambda$ limit,
\begin{equation}
\lim_{\lambda\to\infty}\phi(0)=-\left[\frac{3!}\lambda n(0)\right]^{1/3}\equiv f_\infty(\lambda)\,.\label{eq:phi0 lambda}
\end{equation}
Then to interpolate between the two, we numerically use this functional form
\begin{equation}
\phi(0)\simeq-\left[|f_0|^{-\alpha}+|f_\infty(\lambda)|^{-\alpha}\right]^{-1/\alpha}\,,
\label{eq:phi0 approximation}
\end{equation}
where we find that $\alpha=2$ provides good numerical agreement as shown in fig.~\ref{fig:phi0}.
In particular, we can numerically determine the optimal value of $\alpha$ over an interesting range of $\lambda$ and find that $\alpha\simeq1.97$ at $m=0$ which then increases to $\alpha\simeq2.38$ at $m=1$ and then decreases back down below $2$ as $m$ continues to increase.
This justifies our choice of $\alpha=2$ as a good guess for all cases.

We also note that in the hard sphere case we find the critical value of $\lambda$ that indicates when the transition between the two limits occurs.
\begin{equation}
\lambda_c=\frac{3!}{n^2}\left(\frac{m^2}{1-e^{-mr_s}(1+mr_s)}\right)^3\,,
\end{equation}
which, in the massless limit, becomes
\begin{equation}
\lim_{m\to0}\lambda_c=\frac{3!\times2^3}{n^2r_s^6}\,.
\end{equation}
We see in fig.~\ref{fig:phi0} that eq.~\ref{eq:phi0 approximation} is accurate at the $\lesssim1-2\%$ level for $m=0$ and $\lesssim10\%$ when both $m$ and $\lambda$ are contributing.
This is sufficient to guess the initial value of $\phi(0)$ in the hard shell case and can be used to derive a useful guess for arbitrary density profiles.

\begin{figure}
\centering
\includegraphics[width=\columnwidth]{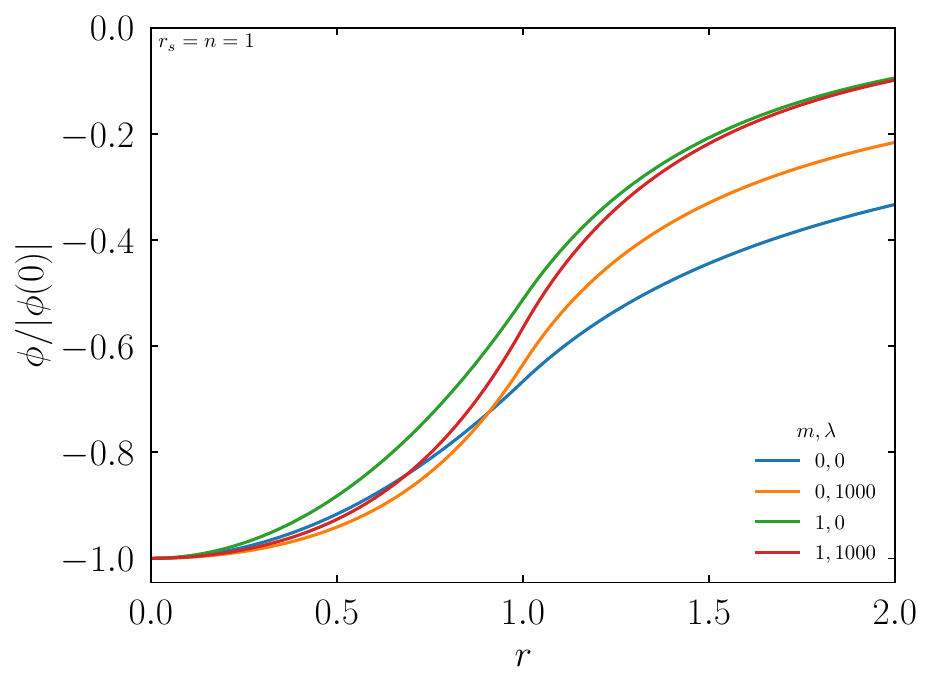}
\caption{The normalized field value for four different pairs of $m$ and $\lambda$ as a function of radius in the hard sphere of charge case.
We have taken the sphere size to be unity $r_s=1$ and the density to also be unity $n=1$.}
\label{fig:phi_m_lambda}
\end{figure}

In fig.~\ref{fig:phi_m_lambda} we show the value of the field as a function of radius for various different values of $m$ and $\lambda$ in the hard sphere of charge case, normalized to the field strength at the origin for easy comparison.
We see that in the case with $m\neq0$ and $\lambda=0$ (Yukawa) the field falls to zero faster than in the $m=\lambda=0$ (electromagnetic) case, as expected due to the $e^{-mr}$ contribution.
The inclusion of the self-interacting $\lambda$ term, however, causes the field to fall off back to zero more slowly at first, and then faster later, leading to a distinct behavior that cannot be recovered with just $m$.
We also see that at large $r$ the $m$ term will dominate over the $\lambda$ term.
This makes sense since as $r$ increases, $|\phi|$ decreases and thus the contribution from $\frac\lambda{3!}\phi^3$ will decrease relative to the $m^2\phi$ term.

\section{Cubic Term}
\label{sec:cubic}
We now briefly comment on the implications of including a $\frac\mu{3!}\phi^3$ term to the potential in eq.~\ref{eq:V} leading to a $\frac\mu2\phi^2$ term in the differential equation in eq.~\ref{eq:KG24 spherical}.
There is likely not an analytic solution to this equation either.
We also notice that, depending on the magnitude and shape of $n(r)$, there may not be a real solution for $\mu>0$ that satisfies the boundary conditions.
For the $\mu<0$ dominating case, at the origin we have that
\begin{equation}
\lim_{\mu\to-\infty}\phi(0)=\sqrt{-\frac{2n(0)}\mu}\,.
\end{equation}
We note that this continues from the same trend of the power dependence of $\phi(0)$ on $n(0)$.
That is, that we expect,
\begin{equation}
\lim_{|c|\to\infty}\phi(0)\propto n^{1/\beta}(0)\,,
\end{equation}
where $c\in\{m,\mu,\lambda,\dots\}$ is the coefficient of each term in the potential $V$ and $\beta$ is the corresponding exponent of $\phi$ in $V'(\phi)$ as it appears in the differential equation.

\section{Code}
We provide code\footnote{The code is available
at \url{https://github.com/PeterDenton/Spherically-Symmetric-SelfInteracting-Scalar-Solver}, see also ref.~\cite{S5}.}
to numerically compute solutions to the Klein-Gordon equation in for arbitrary spherically symmetric source potentials \cite{S5}.
The primary calculations occur in the \verb1S5.cpp1 file.
The code evaluates $\phi(0)$ with the \verb1phi0_KG1 function using the \verb1phi0_Guess1 function as a guess.
Some care may be required to get an appropriate window, depending on the parameters used and the density profile.
Three density profiles are included: \verb1HARD1, \verb1SUN1 from BS05(OP) \cite{Bahcall:2004pz}, and \verb1EARTH1 from the Preliminary Reference Earth Model \cite{Dziewonski:1981xy}.
Then, given $\phi(0)$, the value of the field for a range of radii is calculated with the \verb1KG1 which returns the value of the field as an array.

The \verb1phi0_KG1 function in the code determines the value of $\phi(0)$ such that $\lim_{r\to\infty}\phi(r)=0$.
Specifically, it starts with the guess from eq.~\ref{eq:phi0 approximation} and attempts to root find a function \verb1phi0_KG_helper1 across a window around the initial guess.
The \verb1phi0_KG_helper1 function advances the differential equation in radius and returns a value as soon as a condition is met.
If the derivative turns negative $\phi'<0$ then it returns a negative number that is large in magnitude if it turns negative quickly.
If the value of the field crosses the origin $\phi>0$ then it returns a positive number it is large if it passes the origin quickly.
If neither condition happens after advancing a distance of $100r_s$ then it returns a value related to the value of the field at $100r_s$.

We have checked that this performs as expected whenever analytic expressions are available.

\section{Physics Implications}
Physics models with ultralight bosons have unique and compelling phenomenological and model building effects, see e.g.~\cite{Turner:1983he,Hu:2000ke,Marsh:2015xka,Hui:2016ltb,Davoudiasl:2017jke,Baryakhtar:2017ngi,Kling:2017mif,Kling:2017hjm,Davoudiasl:2019nlo,Banerjee:2019xuy,Zu:2020whs}.
In particular, it has been pointed out that ultralight bosons source by celestial objects can lead to unique dark matter and neutrino signatures either by interacting with the dark matter or neutrinos, or by being the dark matter itself, see e.g.~\cite{Davoudiasl:2018hjw,Smirnov:2019cae,Babu:2019iml,Dev:2020kgz,Dev:2022bae,Davoudiasl:2023uiq,Acevedo:2023owd}.
For example, the ultralight boson could modify the rate at which dark matter is captured in the Earth or the Sun.
Alternatively, the ultralight boson could both be the dark matter and could also be sourced by celestial objects to impact the evolution of neutrinos (active or sterile), axions, or other particles near the object.
In either case the presence of the $\lambda\phi^4$ quartic term would change the physics and is often not considered in the literature.

The presence of the quartic $\lambda\phi^4$ in specific physics models is briefly discussed in \cite{Dev:2022bae} and discussed in more detail in \cite{Davoudiasl:2023uiq}.
For example, even though such models can be largely handled in the semi-classical regime due to the smallness of the couplings that are typically involved, it is pointed out that such a $\lambda\phi^4$ quartic term is generically expected to be present due to loops induced by the coupling of the field to matter, neutrino, or dark matter fields.
Thus unless one chooses to fine tune such a term away or show that it is sufficiently small to not contribute to the dynamics of the field, it is generically expected to be present in any ultralight boson model and must be considered.

Once the quartic self-interaction term is present, it leads to a number of phenomenological effects that are quite different from the simpler Yukawa case with just the $m^2\phi^2$ term.
For example, an ultralight boson dominated by its quartic coupling evolves cosmologically like radiation, until such a time as the quadratic term dominates at which point it evolves like matter \cite{Turner:1983he}.
This has important implications if one considers a scenario where ultralight bosons are the dark matter \cite{Davoudiasl:2023uiq} because it must be matter dominated at least by temperatures of about $T\sim$keV \cite{Das:2020nwc}.

If such a field is sourced by celestial objects due to a coupling to quarks or electrons, scenarios which are of particular relevance for the work presented here, the presence of the $\lambda\phi^4$ may play a very significant role in the resultant phenomenology, in two main ways.
The first is the value of the field at the center of the sphere.
This sets the strength of the potential induced by the scalar field everywhere.
In the regime where the quadratic $m^2\phi^2$ term dominates, we find that
\begin{equation}
\phi(0)\simeq-\frac n{m^2}\left[1-e^{-mr_s}(1+mr_s)\right]\,,
\end{equation}
where $r_s$ is the characteristic radius over which the largest values of $n$ dominate; this is an exact expression for a hard sphere.
For the Earth this approximation would use the average density sourcing the charge in the Earth's core while ignoring the mantle.
For the Sun one could use this to focus on the most dense central region as well and find an excellent approximation, see \cite{Davoudiasl:2023uiq}.
If the characteristic radius $r_s$ is small compared to $1/m$, then this simplifies to $\phi(0)\simeq-\frac n{m^2}$, which is also the solution to eq.~\ref{eq:KG24 spherical} where one sets the derivatives and $\lambda$ terms to zero.

Alternatively, if the $\lambda\phi^4$ term dominates and the $m^2\phi^2$ term can be ignored, we found in eq.~\ref{eq:phi0 lambda} that the field strength in the core takes a very different form:
\begin{equation}
\phi(0)\simeq-\left[\frac{3!}\lambda n(0)\right]^{1/3}\,.
\end{equation}
Not only is the dependence on the model parameters ($m$, $\lambda$) quite different, but more importantly the dependence on the density of the system changes from linear to the $\frac13$ power.
Thus the addition of a $\lambda\phi^4$ term can significantly weaken the strength of the field.

The second phenomenological impact is the shape of the potential sourced by the field as shown in fig.~\ref{fig:phi_m_lambda}
This shows that the behavior of the field in the presence of a quartic term is fundamentally different from that without and could affect the capture rates of dark matter inside celestial objects when ultralight bosons are sourcing a potential especially when multiple scattering and angular effects are in play.
It would also affect the change in the properties of dark matter or neutrinos as a function of radius inside the Earth or the Sun which is potentially probable depending on the parameters and the detection channel \cite{Davoudiasl:2023uiq}.
Finally, it may lead to a dark matter or cosmic neutrino background halo around the Earth which could take a different shape than in the case without the self interaction term.

\section{Conclusions}
In this article we considered the calculation of a scalar field sourced by a spherically symmetric background field with an arbitrary density profile such as a celestial object.
We developed expressions for arbitrary density profiles of the source in the non-self interacting case for a massless or a massive scalar.
In general, however, a scalar field may well have a $\phi^4$ term as well that, as we have shown, acts quite differently from a mass term.
While a closed form expression for the self interacting case with a $\lambda\phi^4$ term seems to not exist, we developed expressions to allow for numerical calculations, provided code to carry out such calculations, and presented some insight into the behavior of these solutions.

\begin{acknowledgments}
The author acknowledges helpful comments from Hooman Davoudiasl and is supported by the United States Department of Energy under Grant Contract No.~DE-SC0012704.
\end{acknowledgments}

\bibliography{KG4}

\end{document}